\documentclass[twocolumn]{emulateapj}
\usepackage{graphicx}
\usepackage{epsfig}

\shorttitle{Dynamical Friction}
\shortauthors{Psaltis}
 
\begin{document}


\title{The Influence of Gas Dynamics on Measuring the Properties of
the Black Hole in the Center of the Milky Way with Stellar Orbits and Pulsars}

\author{Dimitrios Psaltis}
\affil{Astronomy Department,
University of Arizona,
933 N.\ Cherry Ave.,
Tucson, AZ 85721, USA}

\begin{abstract}
Observations of stars and pulsars orbiting the black hole in the
center of the Milky Way offer the potential of measuring not only the
mass of the black hole but also its spin and quadrupole moment,
thereby providing observational verification of the no-hair
theorem. The relativistic effects that will allow us to measure these
higher moments of the gravitational field, however, are very small and
may be masked by drag forces that stars and pulsars experience
orbiting within the hot, tenuous plasma that surrounds the black hole.
The properties of this plasma at large distances from the central
object have been measured using observations of the extended X-ray
emission that surrounds the point source. At distances comparable to
the black-hole event horizon, the properties of the accretion flow
have been constrained using observations of its long-wavelength
emission and polarization, as well as of the size of the emitting
region at 1.3~mm. I use models of the plasma density and temperature
at various distances from the black hole to investigate the effect of
hydrodynamic drag forces on future measurements of the higher moments
of its gravitational field.  I find that hydrodynamic drag does not
preclude measurements of the black hole spin and quadrupole moment
using high-resolution observations of stars and pulsars that orbit
within a few thousand gravitational radii from its horizon.
\end{abstract}


\section{INTRODUCTION}

The black hole in the center of the Milky Way, Sgr~A$^*$ has a number
of characteristics that make it an ideal candidate for testing general
relativity in the strong field regime. It is surrounded by a swarm of
young stars at very close orbits, the tracking of which has already
led to a direct measurement of its mass (see Ghez et al.\ 2008;
Gillessen et al.\ 2009 for recent work). It is accreting from the
surrounding medium at rates high enough to make it a detectable source
both in the X-rays (e.g., Baganoff et al.\ 2003) and at long
wavelengths (e.g., Falcke et al.\ 2008; Yusef-Zadeh et al.\ 2009). It
is also the black hole with a horizon that subtends the largest angle
in the sky, making it possible to directly image the region very close
to its horizon (e.g., Doeleman et al.\ 2008).

Three distinct technological advances in the next decade offer the
possibility of not only improving the accuracy of our knowledge of the
black hole's mass but also of measuring directly its spin and
quadrupole moment. This last quantity of the gravitational field is
not an independent quantity in the Kerr metric; instead it depends in
a very particular way on the mass and spin of the black hole, as
required by the no-hair theorem (see, e.g., Ryan 1995). Measuring all
three moments of the spacetime of Sgr~A$^*$ will, therefore, allow us
to test the black-hole nature of the source, verify the validity of
the no-hair theorem, and consequently test general relativity in the
strong-field regime (see Will 1998; Psaltis \& Johannsen 2011).

Interferometric observations in large telescopes equipped with adaptive
optics (such as GRAVITY on the VLT; Gillessen et al.\ 2010) will
detect and track stars in very close orbits around Sgr~A$^*$. The
orbits of these stars will precess because of the relativistic
corrections to the black-hole's Newtonian field at rates that depend
directly on the compact object's spin and quadrupole moments (Will
2008). 

The discovery of a single radio pulsar in orbit around Sgr~A$^*$ will
also allow for a direct measurement of the black hole's spin and
quadrupole moment, by modeling relativistic effects in the residuals
of its orbital solution (Wex \& Kopeikin 1999). The presence of a
cluster of young, massive stars in orbit around Sgr~A$^*$ makes it
highly likely that such pulsars exist (e.g., Pfahl \& Loeb 2004) and
current searches place a conservative upper limit of $\simeq
90$~pulsars within the central parsec of the galaxy (e.g., Macquart et
al.\ 2010).  The difficulty in detecting these pulsars lies in the
fact that interstellar scattering heavily disperses and broadens the
emission from each pulsar. 

Finally, current mm-VLBI observations of Sgr~A$^*$ with only three
baselines have already confirmed the expectation that the accretion
flow is optically thin at such wavelengths (Doeleman et al.\ 2008) and
has led to the first constraints on the orientation and properties of
the black hole (Broderick et al.\ 2009). Future observations with 5-10
baselines will provide the first direct image of horizon-scale
structures in the vicinity of a black hole (Fish \& Doeleman 2009) and
allow for a model-independent measurement of the black-hole's spin and
quadrupole moment (Johannsen \& Psaltis 2010).

Measuring the mass, spin, and quadrupole moment of the black hole in
the center of the Milky Way with three independent techniques is
crucial in distinguishing relativistic effects from other
astrophysical complications. Indeed, the presence of gas and stars in
the vicinity of the black hole introduce perturbations to the orbits
of stars and pulsars, that may mask or even bias the measurements. The
cluster of stars (known and anticipated) in orbit around Sgr~A$^*$
introduces multipole components to the gravitational field, which in
turn cause precession of their orbits. In order for this classical
effect to be negligible compared to the relativistic precession that
depends on the spin and quadrupole of the black-hole spacetime, the
orbital separations of the stars from the black hole have to be less
than a milliparsec (Merritt et al.\ 2010).

Stars and pulsars orbiting very close to the black hole, however, will
interact with the hot surrounding plasma, which is provided by the
stellar winds of the more distant stars and powers the accretion
luminosity of the black hole. This interaction will cause them to
leave their geodesic orbits and very slowly spiral towards the black
hole. If any of the various components of hydrodynamic drag is
comparable to or larger than the perturbation to the Newtonian
gravitational acceleration due to relativistic effects, then tracking
the orbits of stars or pulsars around the black hole will not lead to
a clean measurement of the moments of its spacetime.

In this article, I first infer the density and temperature of the hot
plasma at different distances from Sgr~A$^*$ using models of the
accretion flow and of its feeding region that are consistent with
current X-ray and millimeter observations. I then estimate the effect
of hydrodynamic drag on the orbits of objects in this plasma and show
that it is negligible for stars and pulsars orbiting within a few
thousand gravitational radii from the black-hole horizon. Through this
work, I will assume that the black hole has a mass of $4.5\times 10^6
M_\odot$ and lies at a distance of $8.4$~kpc, as inferred from the
constrained fit of Ghez et al.\ (2008; see also Yelda et al.\ 2010).

\section{The Density and Temperature of Plasma Around SGR~A$^*$}

The black hole in the center of the Milky Way is believed to be fed by
the winds of massive stars that lie within the central parsec of the
galaxy. A fraction of the mass lost by the stars is gravitationally
captured and accreted by the black hole. Calculating the kinematics
and thermodynamic properties of the plasma within the central plasma
is non-trivial, as the flow is three-dimensional, with little
symmetry, and variable at a large dynamic range of scales.

The feeding of the central black hole has been modeled in the past
using multi-dimensional numerical simulations (e.g., Coker \& Melia
1997; Rockefeller et al.\ 2004; Cuadra et al.\ 2006) or simpler,
one-dimensional analytic models (e.g., Quataert 2004; Shcherbakov \&
Baganoff 2010). Both the analytic and numerical models use similar
prescriptions for the mass-loss rates of the known stars and the newer
works aim to reproduce the X-ray luminosity and temperature of the
extended emission around Sgr~A$^*$ as observed by the Chandra X-ray
Observatory (Baganoff et al.\ 2003). For these reasons, the inferred
density and temperature profiles of the plasma are very similar among
the various models (see, e.g., Fig.~1 of Cuadra et al.\ 2006 for a
comparison). 

\begin{figure}[t]
\psfig{file=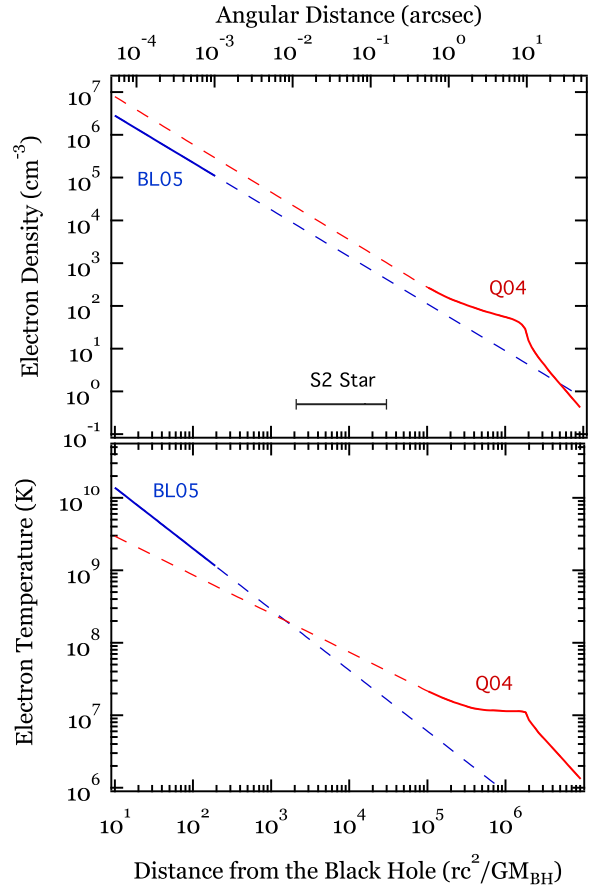,width=3.5in}
\caption{The inferred electron {\em (top)\/} density and {\em
    (bottom)\/} temperature at different distances from Sgr~A$^*$. At
  large distances, the model of Quataert (2004) was designed to fit
  the Chandra observations by Baganoff et al.\ (2003), who inferred
  the electron density and temperature of the extended emission around
  the black hole.  At small distances, the model of Broderick \& Loeb
  (2005) fits the long-wavelength spectrum of the accretion flow, is
  consistent with polarization measurements, and agrees with the size
  of the emitting region at 1.3~mm. The dashed portion of each curve
  represents an extrapolation of each model to distances far beyond
  its intended range of validity.}
\label{fig:density}
\end{figure}

In this work, I will be using the semi-analytic model of Quataert
(2004) as an estimate of the electron density and temperature of the
plasma at distances $10^5-10^7 GM_{\rm BH}/c^2$ from the black hole.
Figure~\ref{fig:density} shows these two properties of the plasma,
both in the region probed by the Chandra observations (solid line) as
well as extrapolated towards much smaller distances from the black
hole (dashed line). The range of orbital separations of the S2 star
from the black hole, which is in the extrapolated regions of the
curves, is also shown for comparison. The orbits of stars for which
relativistic effects can be distinguished over gravitational
perturbations from the stellar cluster itself will lie inside the
orbit of the S2 star, shown with the horizontal solid line (Merritt et
al.\ 2010). As a result, in order to estimate the effect of
hydrodynamic drag on the orbits of nearby stars, I will need to
investigate the validity of the extrapolation of the model towards
small distances.

A second handle on the plasma properties near the black hole horizon
is provided by the millimeter observations of Sgr~A$^*$, which probe
the inner tens of gravitational radii of the accretion flow. During
the last decade, analytic models of two-temperature, radiatively
inefficient accretion flows have successfully accounted for the
mm-to-cm spectra of Sgr~A$^*$ (e.g., Narayan, Yi, \& Mahadevan 1995),
the polarization of its emission (Quataert \& Gruzinov 2000; Agol
2000), as well as the size of its emitting region (\"Ozel, Psaltis, \&
Narayan 2000). 

Broderick \& Loeb (2005; see also Broderick et al.\ 2009, 2011)
constructed a semi-analytic model based on the earlier works and
determined its parameters by fitting simultaneously the latest
mm-to-cm spectra observed from Sgr~A$^*$ (Falcke et al.\ 2008, Marrone
et al,\ 2007), the polarization of its emission (Marrone et
al.\ 2007), and the size of the emitting region at 1.3~mm~(Doeleman et
al.\ 2008). In their model, the electron density $n_e$ and temperature
$T_e$ in the accretion flow scale with radius $r$ as
\begin{equation}
n_e=n_e^0\left(\frac{rc^2}{GM_{\rm BH}}\right)^{-1.1}
\end{equation}
and
\begin{equation}
T_e=T_e^0\left(\frac{rc^2}{GM_{\rm BH}}\right)^{-0.84}\;,
\end{equation}
respectively, where $M_{\rm BH}$ is the mass of the black hole, $G$ is
the gravitational constant, and $c$ is the speed of light. The
best-fit parameters of the model to the observations are
$n_e^0=3.5\times 10^7$~cm$^{-3}$ and $T_e^0=9.5\times 10^{10}$~K
(Broderick et al.\ 2011). Independent MHD numerical simulations of
radiatively inefficient accretion flows, which were compared to the
same data, have reached similar conclusions regarding the overall scale
of the electron density and temperature in the accretion flow
(Mo{\'s}cibrodzka et al.\ 2009; Dexter et al.\ 2010).

Figure~\ref{fig:density} compares the Broderick \& Loeb (2005) model
of the inner accretion flow to the Quataert (2004) model of the
feeding region at large distances from the black hole. Remarkably, the
inferred density of the inner accretion flow is only a factor of $\sim
3$ lower than the extrapolation of the Quataert (2004) model at
distances that are smaller by four orders of magnitude compared to its
intended region of validity (see also similar remarks in Shcherbakov
\& Baganoff 2010). In order to be conservative in calculating the 
effect of hydrodynamic drag on the orbits of objects around Sgr~A$^*$, 
I will use the extrapolation of the Quataert (2004) model throughout
the vicinity of the black hole. Moreover, because in the
regions of interest the electron temperature is in excess of $10^7$~K
(see below) I will assume that the plasma is fully ionized. 

The electron temperature of the inner accretion flow is higher than
the extrapolation of the Quataert (2004) model and has a steeper
radial profile. This is not unexpected, however, as turbulent heating
in the accretion flow will heat the plasma to temperatures higher than
what is calculated in the analytical work, which takes into account
only compressional and adiabatic heating. The discontinuity between
the models is not a serious handicap for this calculation since I will
use the plasma temperature only to infer whether the motion of objects
through the plasma is sub-sonic or super-sonic. In order to achieve
this, however, I will need to make an assumption regarding the ratio
of ion-to-electron temperatures, as the models of both the inner
accretion flow and of the feeding region provide an estimate of only
the electron temperature in the plasma surrounding Sgr~A$^*$.

\begin{figure}
\psfig{file=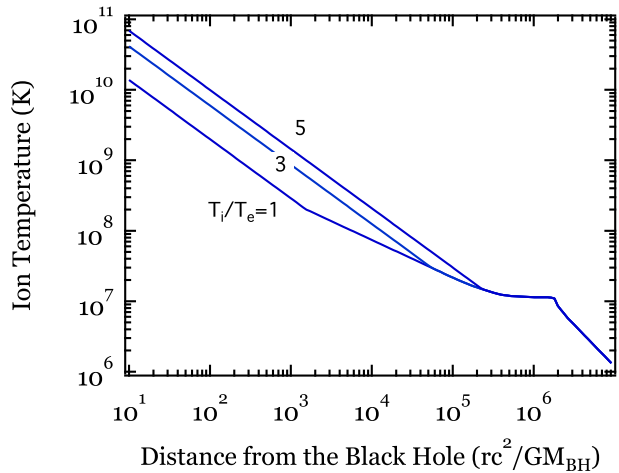,width=3.5in}
\caption{The inferred ion temperature of the accretion flow around
  Sgr~A$^*$. At large distances, it is set equal to the inferred
  electron temperature. At small distances, where the electrons are
  expected to be only weakly coupled with the ions, the ion
  temperature is believed to be $\sim 1-5$ times the electron
  temperature.}
\label{fig:temperature}
\end{figure}

At large distances, ions and electrons are expected to have the same
temperatures. On the other hand, close to the black hole, the
combination of MHD simulations, which follow the dynamics of the ions,
and of radiative transfer calculations, which model the electron
properties, indicate that the ion temperature is $\sim 1-5$ times
the electron temperature (see, e.g., Mo{\'s}cibrodzka et al.\ 2009;
Dexter et al.\ 2010). For the purposes of this work, I will use the
temperature of the model by Quataert (2004) in the outer region of the
plasma and an ion temperature that is $\simeq 1-5$ times the electron
temperature of the Broderick \& Loeb (2005) model in the inner region,
where the latter is higher, as shown in Figure~\ref{fig:temperature}.

\begin{figure}
\psfig{file=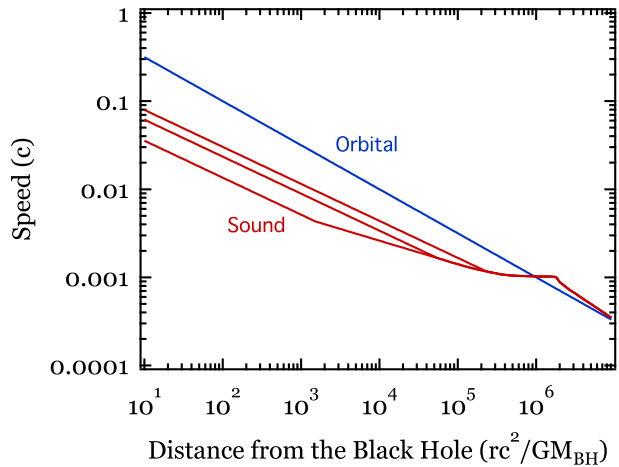,width=3.5in}
\caption{A comparison of the orbital speed to the local sound speed at
  different radii from Sgr~A$^*$. The sound speeds is calculated for
  the inferred electron temperatures shown in
  Figure~\ref{fig:temperature}.  Throughout the inner region of the
  accretion flow, orbital speeds are mildly to highly supersonic.}
\label{fig:speed}
\end{figure}

Figure~\ref{fig:speed} compares the sound speed in the models shown in
Figure~\ref{fig:temperature} to the characteristic orbital speed
$u_{\rm orb}=(GM_{\rm BH}/r)^{1/2}$ at different distances from the
black hole. In all regions of interest, i.e., for distances $\lesssim
10^5 M$, the orbital speed is larger than the local sound speed. As 
a result, the motion of stars and pulsars through the plasma surrounding
Sgr~A$^*$ will be mildly to highly supersonic.

\section{The interaction of Stars and Pulsars with the Hot Plasma}

\begin{figure}[t]
\psfig{file=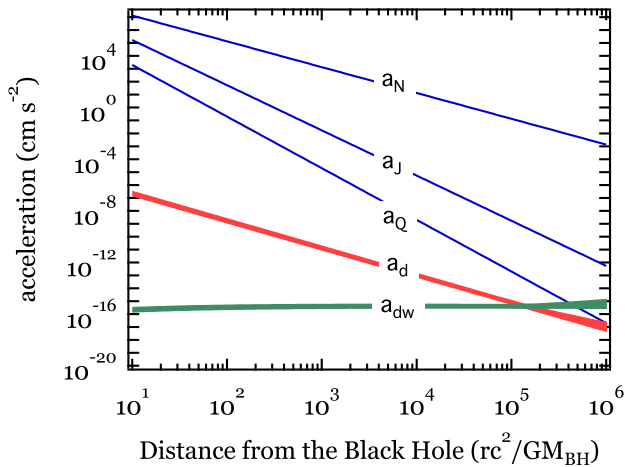,width=3.5in}
\caption{A comparison of various terms of the gravitational
  acceleration experienced by a star of mass $10~M_\odot$ and radius
  $10~R_\odot$ in orbit around Sgr~A$^*$ {\em (blue lines)}, to the
  acceleration due to hydrodynamic drag ($a_{\rm d}$; red line) and
  due to the gravitational interaction of the star with its wake
  ($a_{\rm dw}$; green line). The three blue lines correspond to the
  Newtonian acceleration ($a_{\rm N}$), the term proportional to the
  spin of the black hole ($a_{\rm J}$), and the term proportional to
  its quadrupole moment ($a_{\rm Q}$). The black hole is assumed to be
  described by the Kerr solution and to be mildly spinning
  ($\chi=0.1$). At radii $\lesssim 10^5 M$, hydrodynamic interactions are
  not large enough to preclude a successful measurement of the spin
  and quadrupole moment of the black hole.}
\label{fig:acceleration}
\end{figure}

The motion of a star or a pulsar in the vicinity of Sgr~A$^*$ will be
affected by its interaction with the surrounding plasma to the degree
that it will preclude a clean gravitational experiment, if the
interaction results in accelerations that are comparable to that of
gravity. As a scale against which I will compare the hydrodynamic
accelerations, the Newtonian gravitational acceleration at a distance 
$r$ away from the central black hole is
\begin{eqnarray}
a_{\rm N}&=&\frac{GM_{\rm BH}}{r^2}\nonumber\\ 
&\simeq&1.4\times 10^9 \left(\frac{M_{\rm BH}}{4.5\times 10^6 M_\odot}\right) 
\left(\frac{GM_{\rm BH}}{rc^2}\right)^2~{\rm cm~s}^{-2}\;.
\end{eqnarray}
An observational verification of the no-hair theorem around Sgr~A$*$
will involve measuring the magnitude of the correction to the Newtonian
acceleration that is proportional to the spin and the quadrupole
moment of the black hole and verifying whether these lowest three
multipole moments of its spacetime obey the relation of the Kerr
metric. Corrections to the Newtonian acceleration that depend on the
spin of the black hole are proportional to (see Merritt et al.\ 2010)
\begin{eqnarray}
a_{\rm J}&=&\frac{4\chi G^2 M_{\rm BH}^2}{c^3}\left(\frac{u_{\rm orb}}{r^3}\right)
\nonumber\\
&=&5.4\times 10^9 \chi\left(\frac{M_{\rm BH}}{4.5\times 10^6 M_\odot}\right)^{-1} 
\left(\frac{GM_{\rm BH}}{rc^2}\right)^{7/2}~{\rm cm~s}^{-2}\;,\nonumber\\
\end{eqnarray}
where $\chi$ is the dimensionless spin of the black hole. Finally,
corrections to the Newtonian acceleration that depend on the
quadrupole moment of the spacetime, assuming that the latter is
described by the Kerr solution, are proportional to (Merritt et
al.\ 2010)
\begin{eqnarray}
a_{\rm Q}&=&\frac{3}{2}\chi^2\left(\frac{G^3 M_{\rm BH}^3}{c^2r}\right)^4
\nonumber\\
&=&2.0\times 10^9 \chi^2
\left(\frac{M_{\rm BH}}{4.5\times 10^6 M_\odot}\right)^{-1} 
\left(\frac{GM_{\rm BH}}{rc^2}\right)^{4}~{\rm cm~s}^{-2}\;.\nonumber\\
\end{eqnarray}
Figure~\ref{fig:acceleration} shows the magnitude of the various
contributions to the gravitational acceleration at different 
distances from the central black hole, for an assumed spin of $\chi=0.1$.

An orbiting object in the vicinity of Sgr~A$^*$ interacts with the
surrounding plasma in at least two distinct ways (see also Narayan
2000), which I will now consider in some detail.

First, as the object plows through the plasma, it scatters away the
plasma particles or it gravitationally captures them, depending on its
compactness. In the former case, the orbiting object transfers some of
its linear momentum to the plasma particles, thereby feeling a
hydrodynamic drag. In the latter case, the gravitational focusing of
the trajectories of the plasma particles towards the back side of a
small orbiting object leads to an increase of its linear momentum
(Ruffer 1996). Which of the two cases dominates depends on the ratio
of the Bondi radius to the radius of the object itself.  Given that
the motion of stars and pulsars in the vicinity of Sgr~A$^*$ is
supersonic everywhere in the region of interest, I estimate the Bondi
radius around such objects to be
\begin{eqnarray}
R_{\rm B}&=&\frac{2GM_*}{u_{\rm orb}^2}=\frac{2GM_*}{c^2}
\left(\frac{rc^2}{GM_{\rm BH}}\right)\nonumber\\
&\simeq&4\times 10^{-5}\left(\frac{M_*}{10 M_\odot}\right)
\left(\frac{rc^2}{GM_{\rm BH}}\right)R_\odot\nonumber\\
&\simeq&6\times 10^{5}\left(\frac{M_*}{2 M_\odot}\right)
\left(\frac{rc^2}{GM_{\rm BH}}\right)~{\rm cm}\;.
\end{eqnarray}
In other words, the Bondi radius around a 10~$M_\odot$ star is smaller
than its stellar radius and, therefore, such a star will always feel a
hydrodynamic drag. On the other hand, the Bondi radius around a
neutron star is comparable to its size even at the smallest orbital
separations from the central black hole and rapidly increases for
larger orbits. As a result, Bondi accretion onto the neutron star will
dominate the scattering of the plasma particles and lead to an
increase of its orbital velocity.

I will estimate the magnitude of the acceleration due to the
hydrodynamic drag $a_{\rm d}$ using the relation
\begin{equation}
M_* a_{\rm d}\simeq \pi R_{\rm eff}^2 \rho u_{\rm rel}^2\;,
\label{eq:hydro_drag}
\end{equation}
where $\rho$ is the density of the plasma that I inferred in \S2,
$u_{\rm rel}$ is the relative velocity of the object with respect to
the plasma, which I will set equal to the orbital velocity, and
$R_{\rm eff}$ is the effective radius at which the moving object
interacts with the plasma. For a normal star, the Bondi radius
is much smaller than the stellar radius. I will, therefore, set 
the effective radius equal to the stellar radius $R_*$ and obtain
\begin{eqnarray}
a_{\rm d}&\simeq& 1.1\times 10^{-9}
\left(\frac{M_*}{10 M_\odot}\right)^{-1}
\left(\frac{R_*}{10 R_\odot}\right)^{2}\nonumber\\
&&\qquad\qquad
\left(\frac{n_e}{10^4~{\rm cm}^{-3}}\right)
\left(\frac{GM_{\rm BH}}{rc^2}\right)~{\rm cm~s}^{-2}\;,
\end{eqnarray}
where I have assumed a cosmic abundance of fully ionized material to
express the mass density in terms of an electron density. Because the
electron density in the accretion flow scales as $n_e\simeq r^{-1}$
(see \S2), the acceleration due to the hydrodynamic drag scales as
$r^{-2}$ and, therefore, keeps an almost constant ratio with the
Newtonian acceleration. Figure~\ref{fig:acceleration} compares the
acceleration due to hydrodynamic drag to the various terms of the
gravitational acceleration. Clearly, hydrodynamic drag on a
10~$M_\odot$ star introduces only negligible perturbations to even the
quadrupole term of the gravitational acceleration in all regions of
interest. Note, however, that in the above estimates, I assumed a
radius of the star that is typical for the main sequence. If the star
is a giant, the acceleration due to the hydrodynamic drag will be
significantly larger.

\begin{figure}[t]
\psfig{file=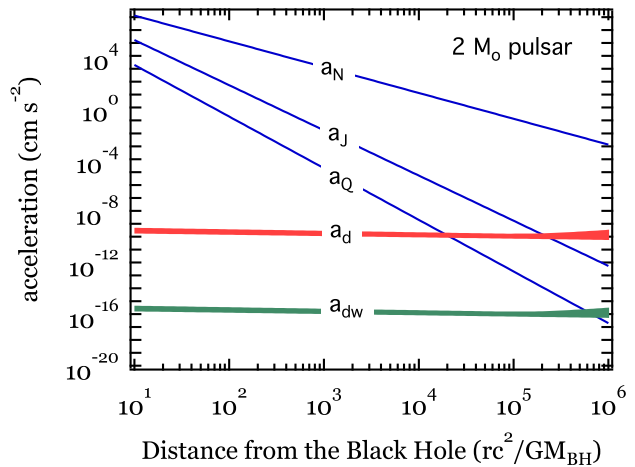,width=3.5in}
\caption{Same as in Figure~\ref{fig:acceleration} for a 2~$M_\odot$
  neutron star. At radii $\lesssim 10^4 M$, hydrodynamic interactions
  are not large enough to preclude a successful measurement of the
  spin and quadrupole moment of the black hole.}
\label{fig:acceleration_psr}
\end{figure}

The hydrodynamic drag on a pulsar has a different dependence on the
distance from the black hole compared to that of a 10~$M_\odot$
star. The effective radius in this case is the Bondi radius, which
increases linearly with increasing distance from the black
hole. Setting $R_{\rm eff}=R_{\rm B}$ in
equation~(\ref{eq:hydro_drag}), I obtain, for a 2~$M_\odot$ neutron
star,
\begin{eqnarray}
a_{\rm d}&\simeq& 6.2\times 10^{-14}
\left(\frac{M_*}{2 M_\odot}\right)^{-1}\nonumber\\
&&\qquad\qquad
\left(\frac{n_e}{10^4~{\rm cm}^{-3}}\right)
\left(\frac{rc^2}{GM_{\rm BH}}\right)~{\rm cm~s}^{-2}\;,
\end{eqnarray}
Figure~\ref{fig:acceleration_psr} compares the hydrodynamic drag
exerted on the neutron star to the various terms of the gravitational
acceleration. Because the electron density in the region of interest
scales as $n_e\sim r^{-1}$, the acceleration due to the hydrodynamic
drag on the neutron star is nearly independent of the distance from
the black hole. It becomes comparable to the quadrupole contribution
to the gravitational acceleration at $\simeq 10^4$ gravitational
radii.

As an object moves through the plasma in the vicinity of the black
hole, it will also generate a wake behind it. The gravitational
interaction between the object and the wake will result in an
additional drag on the object, which will decelerate its motion. The
magnitude of this effect has been calculated by Ostriker (1999) for
the uniform motion of a star in a medium of constant density and more
recently by Kim \& Kim (2007, 2009) and Kim (2010) for the case of a
circular motion in a stratified medium. The results of these papers
were used by Narayan (2000) and by Barausse (2007) and Barausse \&
Rezolla (2008) to estimate the effect of hydrodynamic drag on the
properties of gravitational waves emitted during extreme mass-ratio
inspirals. 

Following these works, I estimate the magnitude of the resulting
acceleration using the expression
\begin{equation}
M_* a_{\rm dw}\simeq 4\pi \ln\left(\frac{R_{\rm max}}{R_{\rm min}}\right)
\left(\frac{GM_*}{u_{\rm rel}^2}\right)\rho\;. 
\end{equation}
Note that the drag due to the gravitational interaction with the 
wake decreases with increasing relative velocity, because as the
velocity increases, the opening angle of the wake decreases.

For a 10~$M_\odot$ star in orbit at radius $r$ around the central
black hole, the appropriate values for the maximum and minimum 
distances of the wake are $R_{\rm max}\simeq r$ and $R_{\rm min}=R_*$,
respectively. Setting, as above, the relative velocity equal to the
orbital velocity $u_{\rm orb}$, I obtain
\begin{eqnarray}
a_{dw}&=&9\times 10^{-15}\left\{
\ln\left[\left(\frac{rc^2}{GM_{\rm BH}}\right) \left(\frac{10
    R_\odot}{R_*}\right)\right]-0.046\right\}\nonumber\\ && \left(\frac{M_{\rm
    BH}}{4.5\times 10^6 M_\odot}\right) \left(\frac{n_e}{10^4~{\rm
    cm}^{-3}}\right) \left(\frac{rc^2}{GM_{\rm BH}}\right)~{\rm
  cm~s}^{-2}\;.\nonumber
\end{eqnarray} 
Note that, because the electron density scales as $n_e\sim r^{-1}$,
this acceleration is also nearly constant throughout the region of
interest.  Figure~\ref{fig:acceleration} compares the acceleration due
to the interaction with the wake of a 10~$M_\odot$ to the
gravitational acceleration from the central black hole. Within $\simeq
10^6$ gravitational radii from the central black hole, this component of
the hydrodynamic drag is negligible compared to the quadrupole
component of the gravitational acceleration.

\mbox{}\\

For the case of a 2~$M_\odot$ neutron star, the minimum distance of the
wake is the Bondi radius, i.e., $R_{\rm min}=R_{\rm B}$ and the acceleration
due to the gravitational interaction of the star with its wake is
\begin{eqnarray}
a_{dw}&=&4.1\times 10^{-21}\nonumber\\
&&\left\{
\ln\left[\left(\frac{M_{\rm BH}}{4.5\times 10^6 M_\odot}\right)
  \left(\frac{M_*}{2 M_\odot}\right)^{-1}\right]
+13.9\right\}\nonumber\\ 
&& \left(\frac{M_*}{2 M_\odot}\right) 
\left(\frac{n_e}{10^4~{\rm
    cm}^{-3}}\right) \left(\frac{rc^2}{GM_{\rm BH}}\right)~{\rm
  cm~s}^{-2}\;.\nonumber\\
\end{eqnarray} 
As before, the the acceleration due to the gravitational interaction
of the neutron star with its wake depends only weakly on its distance
from the black hole. Figure~\ref{fig:acceleration_psr} compares the
magnitude of this acceleration to the gravitational accelerations from
the central black hole demonstrating that the former is again
negligible in all regions of interest.

\newpage

\section{Conclusions}

In this paper, I used the most recent models of the X-ray and
millimeter observations of the black hole in the center of the Milky
Way in order to estimate the density and temperature of the plasma in
a wide range of distances from the central black hole. I then calculated
the effect of hydrodynamic drag on the orbits of star and pulsars that
can be used to probe relativistic effects and test the no-hair
theorem. 

I found that in both the case of orbiting stars and of pulsars, the
hydrodynamic drag dominates over the gravitational interaction of the
object with its wake. For the case of 10~$M_\odot$ stars or
2~$M_\odot$ pulsars around Sgr~A$^*$, the hydrodynamic drag is
negligible compared to the quadrupole terms of the gravitational
acceleration, as long as their orbits are within $\simeq 10^5$ and
$\simeq 10^4$ gravitational radii, respectively, from the central
black hole. Albeit tight, such orbits are also required for the
perturbations in the gravitational field due to the presence of the
stellar cluster to be negligible compared to relativistic effects
(Merritt et al.\ 2010). In other words, for those stars and pulsars
for which the interactions with the other objects of the stellar
cluster are negligible, hydrodynamic drag will not preclude measuring
the spin and quadrupole moment of the black hole spacetime.

\acknowledgements

I thank Stefan Gillessen, Tim Johannsen, and Feryal \"Ozel for useful
comments on the manuscript. This work was supported by the NSF CAREER
award NSF 0746549.

\end{document}